%% file: blife.tex
\documentclass[11pt]{article}
\usepackage{graphicx}

\newcommand{\BABARPubYear}    {00}

\newcommand{\BABARProcNumber} {21}
\newcommand{\SLACPubNumber} {8678}

\input babarsym
\long\def\inst#1{\par\nobreak\kern 4pt\nobreak
    {\it #1}\par\vskip 10pt plus 3pt minus 3pt}

\begin{document}
{\pagestyle{empty}

\begin{flushright}
SLAC-PUB-\SLACPubNumber \\
\babar-PROC-\BABARPubYear/\BABARProcNumber \\
October, 2000 \\
\end{flushright}

\par\vskip 4cm

\begin{center}
\Large \bf B lifetime measurements with exclusively reconstructed B decays
\end{center}
\bigskip

\begin{center}
\large 
Chih-hsiang Cheng\\
Department of Physics, Stanford University \\
Stanford, California 94305, USA \\
(for the \lbabar\  Collaboration)
\end{center}
\bigskip \bigskip

\begin{center}
\large \bf Abstract
\end{center}

Data collected with the \fbabar\  detector at the PEP-II asymmetric
B Factory at SLAC are used to measure the lifetimes of the \Bz\  and \Bu\
mesons. The data sample consists of 8.3~\invfb\  collected near the
\FourS\  resonance. \Bz\  and \Bu\  mesons are fully reconstructed in
several exclusive hadronic decay modes to charm and charmonium final
states. The \B\  lifetimes are determined from the flight length
difference between the two \B\  mesons.
The preliminary measurements of the lifetimes are
\begin{eqnarray*}
\tau_{\Bz} &=& 1.506\pm 0.052\ {\rm (stat)} \pm 0.029\ {\rm (syst)}\ \ps, \\
\tau_{\Bu} &=& 1.602\pm 0.049\ {\rm (stat)} \pm 0.035\ {\rm (syst)}\ \ps,
\end{eqnarray*}
and of their ratio is
$$ \tau_{\Bu }/\tau_{\Bz }=1.065\pm 0.044 \ {\rm (stat)} \pm 0.021 \ {\rm(syst)}.$$
}

\vfill
\begin{center}
Contributed to the Proceedings of the DPF 2000 Meeting \\
of the Division of Particles and Fields of the American Physical Society,\\
8/9/2000---8/12/2000, Columbus, Ohio
\end{center}

\vspace{1.0cm}
\begin{center}
{\em Stanford Linear Accelerator Center, Stanford University, 
Stanford, CA 94309} \\ \vspace{0.1cm}\hrule\vspace{0.1cm}
Work supported in part by Department of Energy contract DE-AC03-76SF00515.
\end{center}

\setcounter{footnote}{0}
\renewcommand{\thefootnote}{\alph{footnote}}

\section{Introduction}
\noindent

This paper presents preliminary measurements of the \Bpm\  and \Bz/\Bzb\
lifetimes and their ratio with data collected by the \babar\ 
detector\cite{bib:TDR}
at the PEP-II asymmetric-energy \B\  Factory. At the \FourS, \B\  mesons are
produced in \Bu\Bub\  or \Bz\Bzb\  pairs. In PEP-II, the center-of-mass
frame is boosted roughly along the $z$ axis\footnote{The axis of the
PEP-II beams is tilted by 20 mrad with respect to the \fbabar\ $z$
axis.} \
($\beta\gamma\simeq0.56$).
In this analysis, one of the \B\  mesons is fully reconstructed in a
variety of clean hadronic decay
modes to charm and charmonium final states. An inclusive technique is
performed to reconstruct the decay vertex of the opposite \B\  meson, and
the lifetime is determined from the distance along the $z$-axis between
the decay vertices of the two \B\  mesons 
($\Delta z=z_{\rm rec}-z_{\rm opp}$).
The true value of $\Delta z$ is distributed exponentially with
an average of $\langle |\Delta z| \rangle= (\beta\gamma)^z_\B c
\tau_B\simeq(p^z_{\FourS}/m_{\FourS})c\tau_B$.

\section{Event sample and vertex reconstruction}
\noindent
The data used in this analysis were collected by the \babar\  detector
at the PEP-II storage ring in the period from January to June
2000. The total luminosity is 7.4\invfb\  collected at the \FourS\
resonance and 0.9\invfb\ 40\mev\  below the resonance. The number of
produced \BB\  pairs is estimated to be $8.4\times 10^6$.

\Bz\  and \Bu\  mesons are reconstructed in the following
modes (and their charge conjugates):
\BztoDdstpi, $\Ddstm\rho^+$,
$\Ddstm a_1^+$, $\jpsi\Kstarz$ and \ButoDdstpi, $\jpsi K^+$, 
$\psitwos K^+$. All final state particles are reconstructed. 
The signal region for each decay mode in the selected sample is defined
by the three standard deviation bands in 
the two-dimensional distribution of the kinematic
variables, $\Delta E$ and $m_{\rm ES}$\footnote{$\Delta E=
E^*_{rec}-E^*_b$, $m_{\rm ES}=\sqrt{E^{*2}_b-{\mbox{\boldmath
$p$}_{\rm rec}^{*2}}}$ where $E^*_{rec}$ and ${\mbox{\boldmath $p$}_{\rm
rec}^*}$ are the measured \B\  candidate energy and momentum, and
$E^*_b$  is the beam energy, all defined in the center-of-mass
frame.}.
The resolution on $m_{\rm ES}$ is about 3\mevcc,
and that on $\Delta E$ varies from
mode to mode between 12 and 40 \mev. 
The purities of final samples are approximately 90\% (Fig.1).

A geometric and kinematic fit of the fully reconstructed $\B_{\rm rec}$
is performed and is required to converge. 
The vertex resolution ranges from 45
to 65 \mum, depending on the mode.
The vertex of the opposite \B\  is determined using all the tracks that
are not associated with the $\B_{\rm rec}$, along with the $\B_{\rm
opp}$ ``pseudotrack'', which is estimated
from the momentum difference between  \FourS\  and $\B_{\rm rec}$, and the beamspot. 
The fitting procedure is
repeated, after removing tracks and ${\rm V^0}$ decays that result in
poor fits, until no track contributes more than a certain value to $\chi^2$. 
This algorithm gives one standard deviation errors of 115\mum\  
and biases around 25\mum\  with Monte Carlo simulation.
The Monte Carlo simulation is also used to study the $\Delta z$ resolution
function of the residual $\delta(\Delta z)=
(\Delta z)_{\rm measured}-(\Delta z)_{\rm generated}$ and the pull
$\delta(\Delta z)/\sigma(\Delta z)$. Two-Gaussian fits successfully
describe the Monte Carlo residual and pull distributions. They give a
one standard deviation width of 130\mum\  and a bias of 24\mum\  for
the residual. Because it is dominated by the
$\B_{\rm opp}$ vertex, the resolution function shape is essentially
the same for all modes we have considered. 

\section{Lifetime fits and systematic uncertainties}
\noindent
The lifetime is extracted from the $\Delta z$ distribution of the
selected events with an unbinned maximum likelihood fit. The
measurements performed on each event~$i$ are 
represented by three input numbers:
$(\Delta z)_i$, its error $\sigma_i$, and 
the probability for an event~$i$ to come from signal based on the
$m_{\rm ES}$ spectrum.
All events that satisfy $5.2 < m_{\rm ES} < 5.3$\gevcc\  and have $\Delta E$ in
the signal region are input to the fit. 
The fit determines the \B\  lifetime, the proportion of outliers,
and two sets of parameters 
that describe the resolution function and the $\Delta z$ shape of the background.
The signal distribution  
is represented by a convolution of the
theoretical $\Delta z$ distribution (two exponential wings) and the
pull representation of the $\Delta z$ resolution function. 
A function of the form $G\otimes(1+E)$ is used\footnote{The sum of an unbiased
Gaussian and the convolution of the same Gaussian with a decaying
exponential.  This resolution function leads to the smallest overall error
with present statistics. }.
The background $\Delta z$ distribution 
is described by the sum of a single Gaussian and two independent
exponential tails, one for positive $\Delta z$ and one for negative
$\Delta z$.
The outliers are described by a wide symmetric Gaussian with
a fixed width of 2500\mum.

We fit the $\Delta z$ distributions of \Bz\  and \Bu\  samples
simultaneously with a single set of parameters for the resolution
function but different sets of parameters (one per charge) to
describe the lifetime, the background and the outliers (Fig.1).
The lifetime ratio is determined with a similar combined fit with the
\Bz\  lifetime and the ratio being free parameters. The statistical
errors are about 3.5\%. 
About 2\% of the statistical error is due to the high correlation
between lifetime and the $\Delta z$ resolution function parameters.

The dominant systematic uncertainties are the limited Monte Carlo
statistics for checking $\Delta z$ distortion due to event selection,
the modeling of $\Delta z$ outliers, the length scale measured
along $z$, and the background modeling. Each contributes about 1\% to the 
systematic uncertainty. All other systematic uncertainties we have studied,
including the choice of parameterization for the resolution function  and
errors on the boost measurements, are less than 0.5\%. The approximation of using
the \FourS\  boost for both \B\  mesons in an event results in a 0.4\% shift.
The central values of the lifetimes are corrected for this shift.

\begin{figure}[htbp]
\vspace*{-20pt}
\begin{center}
\begin{tabular}{lr}
\mbox{\includegraphics[width=2.7in]{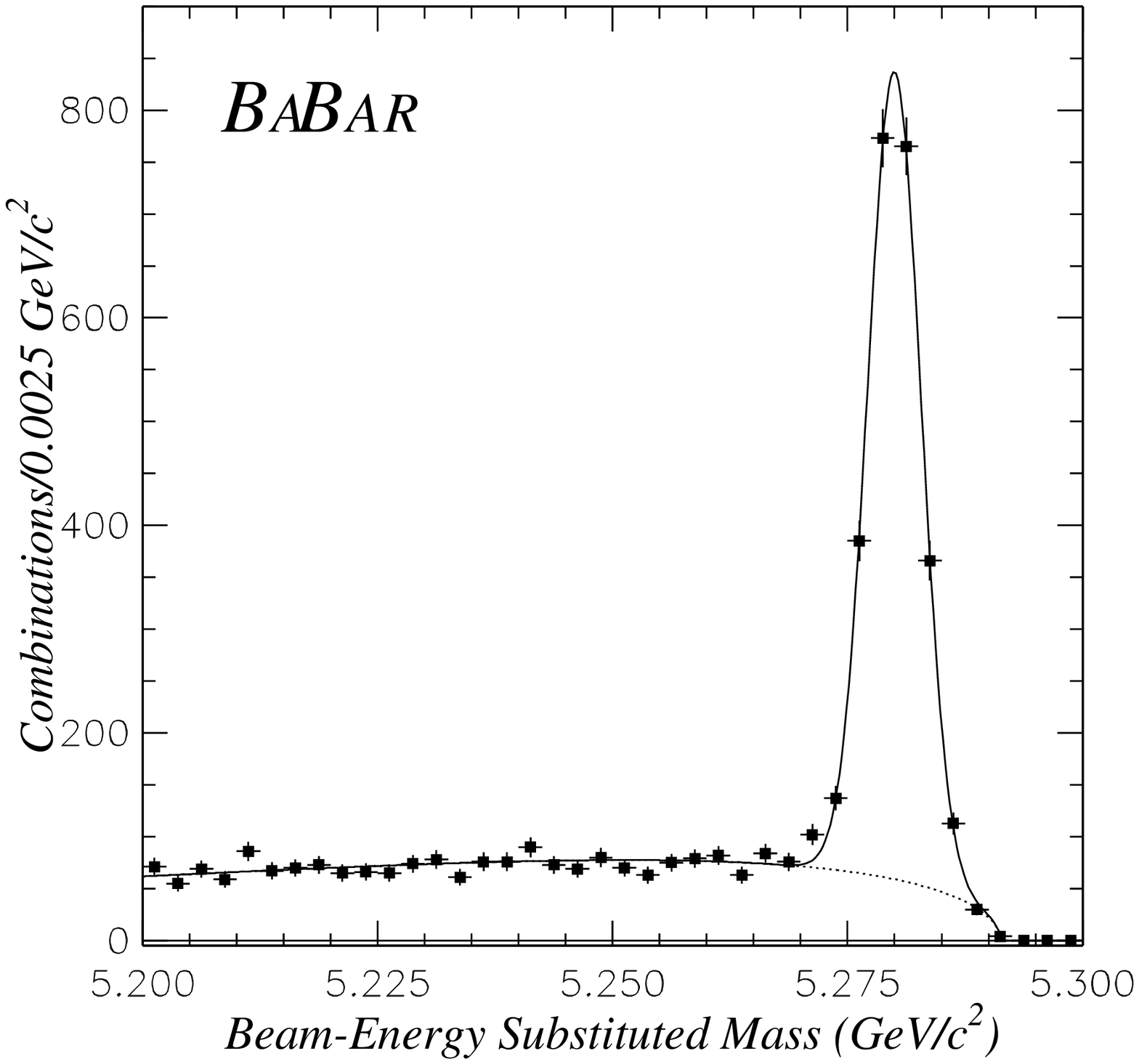}} &
\mbox{\includegraphics[width=2.7in]{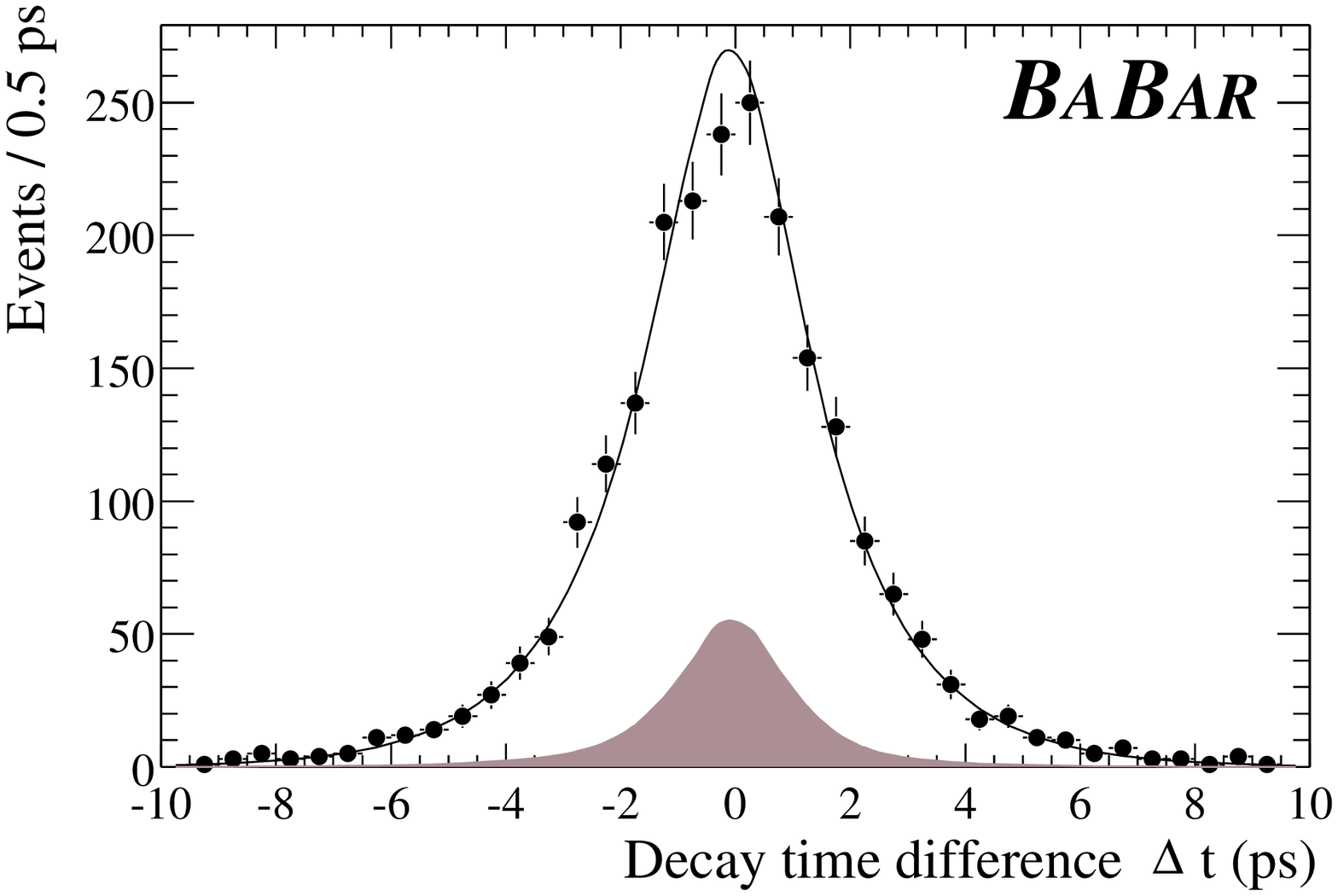}} \\
\end{tabular}
\vspace*{13pt}
\caption{Left: $m_{\rm ES}$ distribution for all the hadronic modes
for \Bz, fit to the sum of a Gaussian distribution and 
the ARGUS background parameterization\cite{bib:argusfunction}. The total
yields 
in all \Bz\  and \Bu\  modes are $2210\pm58$ 
and $2261\pm53$ (not shown), respectively. Right: $\Delta t$
distribution for \Bz/\Bzb. The
result of the lifetime fit is superimposed. The background is shown by
the hatched distribution.}
\end{center}
\label{fig:breco}
\end{figure} 

The preliminary results for the \B\  meson lifetimes are
$$\tau_{\Bz} = 1.506\pm 0.052  \pm 0.029\ \ps, \ \ \ \ \
\tau_{\Bu} = 1.602\pm 0.049  \pm 0.035\ \ps,  $$
and for their ratio is
$$ \tau_{\Bu }/\tau_{\Bz }=1.065\pm 0.044 \  \pm 0.021.$$
The first error is statistical and the second is systematic.
The results are consistent with previous \B\  lifetime
measurements\cite{bib:PDG2000}  
and competitive with the most precise ones.

\end{document}

%% file: babarsym.tex
\def\lbabar{\mbox{{\large\sl B}\hspace{-0.4em} {\normalsize\sl A}\hspace{-0.03em}{\large\sl B}\hspace{-0.4em} {\normalsize\sl A\hspace{-0.02em}R}}}
\def\babar{\mbox{\normalsize\sl B\hspace{-0.4em} {\small\sl A}\hspace{-0.37em} \normalsize\sl B\hspace{-0.4em} {\small\sl A\hspace{-0.02em}R}}}

\def\fbabar{\mbox{\footnotesize\sl B\hspace{-0.4em} {\scriptsize\sl A\hspace{-0.4em} {\footnotesize\sl B\hspace{-0.4em} {\scriptsize\sl A\hspace{-0.1em}R}}}}}

\def\pip   {\ensuremath{\pi^+}}

\def\Kbar  {\ensuremath{\kern 0.2em\overline{\kern -0.2em K}}}

\def\Kstarz  {\ensuremath{K^{*0}}}

\def\Kstarb   {\ensuremath{{\Kbar}\kern0.03em {\raise.18ex\hbox{$^*$}}}}

\def\Kzb   {\ensuremath{{\Kbar}\kern0.03em {\raise.1ex\hbox{$^0$}}}}
\def\KzKzb {\ensuremath{K^0 {\kern -0.16em \Kzb}}}

\def\Dbar  {\ensuremath{\kern 0.2em\overline{\kern -0.2em D}}}
\def\Dzb   {\ensuremath{{\Dbar}\kern0.03em {\raise.3ex\hbox{$^0$}}}}
\def\DzDzb {\ensuremath{D^0 {\kern -0.16em \Dzb}}}

\def\Dstarb   {\ensuremath{{\Dbar}\kern0.03em {\raise.18ex\hbox{$^*$}}}}

\def\Ddstbar  {\ensuremath{\Dbar^{(*)0}}}

\def\Ddstm    {\ensuremath{D^{(*)-}}}

\def\Bz    {\ensuremath{B^0}}
\def\Bbar  {\ensuremath{\kern 0.18em\overline{\kern -0.18em B}}}
\def\Bzb   {\ensuremath{{\Bbar}\kern0.03em {\raise.3ex\hbox{$^0$}}}}
\def\Bu    {\ensuremath{B^+}}
\def\Bub   {\ensuremath{B^-}}
\def\Bpm   {\ensuremath{B^\pm}}

\def\BB    {\ensuremath{B\Bbar}} 
\def\BzBzb {\ensuremath{B^0 {\kern -0.16em \Bzb}}}
\def\B	   {\ensuremath{B}}

\def\jpsi  {\ensuremath{{J\mskip -3mu/\mskip -2mu\psi\mskip 2mu}}} 
\def\psitwos {\ensuremath{\psi{\rm (2S)}}}
\mathchardef\Upsilon="7107
\def\Y#1S{\ensuremath{\Upsilon{\rm(#1S)}}}

\def\FourS {\Y4S}

\mathchardef\Delta="7101
\mathchardef\Xi="7104
\mathchardef\Lambda="7103
\mathchardef\Sigma="7106
\mathchardef\Omega="710A
\def\Deltabar   {\ensuremath{\kern 0.25em\overline{\kern -0.25em \Delta}}{}}
\def\Lbar {\ensuremath{\kern 0.2em\overline{\kern -0.2em\Lambda\kern 0.05em}\kern-0.05em}{}}
\def\Sigbar{\ensuremath{\kern 0.2em\overline{\kern -0.2em \Sigma}}{}}
\def\Xibar{\ensuremath{\kern 0.2em\overline{\kern -0.2em \Xi}}{}}
\def\Obar{\ensuremath{\kern 0.2em\overline{\kern -0.2em \Omega}}{}}
\def\Nbar{\ensuremath{\kern 0.2em\overline{\kern -0.2em N}}{}}

\def\BztoDdstpi {\ensuremath{\Bz \to \Ddstm \pip}}
\def\ButoDdstpi {\ensuremath{\Bu \to \Ddstbar \pip}}

\def\ev   {\ensuremath{\rm \,e\kern -0.08em V}}
\def\kev  {\ensuremath{\rm \,ke\kern -0.08em V}} 
\def\mev  {\ensuremath{\rm \,Me\kern -0.08em V}} 
\def\gev  {\ensuremath{\rm \,Ge\kern -0.08em V}} 
\def\gevc {\ensuremath{{\rm \,Ge\kern -0.08em V\!/}c}} 
\def\tev  {\ensuremath{\rm \,Te\kern -0.08em V}}
\def\mevc {\ensuremath{{\rm \,Me\kern -0.08em V\!/}c}} 
\def\gevcc{\ensuremath{{\rm \,Ge\kern -0.08em V\!/}c^2}} 
\def\mevcc{\ensuremath{{\rm \,Me\kern -0.08em V\!/}c^2}}

\def\mum  {\ensuremath{\,\mu\rm m}} 

\def\invfb   {\ensuremath{\mbox{\,fb}^{-1}}}
\def\mus  {\ensuremath{\rm \,\mus}}

\def\ps   {\ensuremath{\rm \,ps}}

\def\mus        {\ensuremath{\,\mu{\rm s}}}    
\def\ps         {\ensuremath{{\rm \,ps}}}   
\def\gsim{{~\raise.15em\hbox{$>$}\kern-.85em
          \lower.35em\hbox{$\sim$}~}}
\def\lsim{{~\raise.15em\hbox{$<$}\kern-.85em
          \lower.35em\hbox{$\sim$}~}}

\def\to                 {\ensuremath{\rightarrow}}

\def\pep2{PEP-II}

\def\chic1{\ensuremath{\chi_{c1}}}
\def\chic2{\ensuremath{\chi_{c2}}}
\def\chic3{\ensuremath{\chi_{c3}}}

\newcommand{\eqref}[1]{Eq.~(\ref{eq:#1})}

\def\jetset74   {\mbox{\tt Jetset \hspace{-0.5em}7.\hspace{-0.2em}4}}